\documentclass[final,5p,times,preprint]{elsarticle}
\usepackage[utf8]{inputenc}
\usepackage{amsfonts}
\usepackage{amsmath}
\usepackage{amssymb}
\usepackage{bm}
\usepackage{lineno,hyperref}
\usepackage{comment}
\usepackage{xcolor}
\usepackage{booktabs}
\usepackage{multirow}
\modulolinenumbers[5]
\usepackage[normalem]{ulem}

\journal{Physics Letters B}

\biboptions{comma,sort&compress}

\usepackage[normalem]{ulem} 
\def\nuc#1#2{\relax\ifmmode{}^{#1}{\protect\text{#2}}\else${}^{#1}$#2\fi}

\begin{document}

\begin{frontmatter}

\title{ Constraining the N=16 Shell Gap in $^{17}$C via Transfer to the Continuum in the $^{16}\text{C}(d,p)^{17}\text{C}$ Reaction
%
%
}

\author[FAMN]{P. Punta}
\cortext[mail]{Corresponding author}
\ead{ppunta@us.es}

\author[FAMN]{J. A. Lay}
\author[FAMN]{A. M. Moro}
\author[USC]{J. Lois-Fuentes}
\author[USC]{B. Fern\'andez-Dom\'inguez}

\address[FAMN]{Departamento de F\'{\i}sica At\'omica, Molecular y Nuclear, Facultad de F\'{\i}sica, Universidad de Sevilla, Apartado 1065, E-41080 Sevilla, Spain}
\address[USC]{IGFAE and Departamento de F\'isica de Part\'iculas, Universidad de Santiago de Compostela, 15758, Santiago de Compostela, Spain}

\begin{abstract}

Recently, a semi-microscopic structure model has been presented to study the structure of a weakly-bound, two-body nucleus with a deformed core,  including Pauli-blocking effects. The model has been successfully applied within the adiabatic distorted wave approximation (ADWA) reaction framework to study the reactions $^{16}$C(d, p)$^{17}$C, restricting the analysis to bound states of the residual $^{17}$C nucleus. In these calculations, the structure of $^{17}$C is described using the recently presented semimicroscopic Nilsson+AMD model (NAMD), considering different Pauli-blocking methods. In the present work, the analysis is extended to  unbound states of this nucleus with the aim of constraining the location of the $1d_{3/2}$ single-particle strength and infer the $N=16$ shell-gap.   Comparing the measured energy differential cross section for this reaction with calculations in which the position of the  $1d_{3/2}$ orbital is arbitrarily varied, we conclude that a large shell-gap ($>$5~MeV) is required, in agreement with recently reported value from [J. Lois-Fuentes \emph{et al.}, Phys. Lett. B 867, 139600 (2025)].

\end{abstract}


\end{frontmatter}


\section{Introduction}
The study of direct reactions involving weakly bound exotic nuclei is an active field in nuclear physics that has been boosted by the continuous development of radioactive beam facilities.
Some exotic nuclei have a rather different ratio of protons to neutrons from that of stable nuclei, which confers them unique properties.
For example, we can find shell-closures that do not fit the well-known magic numbers for stable nuclei.
During the last few decades, the evolution of the nuclear shell structure far from the stability line has been the focus of much attention\cite{Otsuka20}. 

For neutrons in the $s$-$d$-shell the appearance of the $N=16$ shell closure and disappearance of the $N=20$ near the neutron dripline for Oxygen is of particular interest~\cite{Hoffman08,Tshoo12}.
The large $N=16$ shell gap is expected to be maintained for neutron-rich carbon isotopes, and recent experimental measurements from GANIL suggest that this is the case for $^{16-17}$C~\cite{LOISFUENTES25}.
With the aim of locating the $1d_{3/2}$ orbital, a transfer reaction $^{16}\text{C}(d,p)^{17}\text{C}$ was performed populating unbound states of the \nuc{17}{C} system.
Although the initial analysis of these experimental data supports the hypothesis, it is important to compare the data with theoretical reaction calculations that adequately describe the structure of $^{17}$C, with the challenge of properly dealing with the continuum of a nucleus where the deformation plays also a relevant role.

$^{17}$C is a good example of exotic nucleus composed of a weakly bound neutron and a deformed core.
This system has already been described using deformed two-body models showing promising results~\cite{Punta23,Punta25}.
These models have been successfully applied to the study of the reaction $^{16}\text{C}(d,p)^{17}\text{C}$ populating bound states of $^{17}$C.
The results of the theoretical calculations nicely agree with the experimental data from GANIL~\cite{Pereira}, especially when Pauli blocking effects are taken into account~\cite{Punta25}.
These effects are associated with the blocking of single-particle states occupied by the core nucleons, since the factorisation of these two-body models does not allow complete antisymmetrisation of the wave function.

In this work, we extend the application of the developed deformed two-body models to the study of transfer reactions but populating unbound states in the continuum.
To do this, we have implemented in the transfer-to-the-continuum calculations the pseudo-states discretisation method successfully applied in coupled channels calculations for breakup reactions~\cite{deD14,deD17}.
Therefore, we present a practical tool for the theoretical study of transfer to the continuum reactions, which aligns with present experimental activities aimed at elucidating the evolution of single-particle states in weakly bound and even unbound nuclei.
Then, this formalism is applied to study the reaction $^{16}\text{C}(d,p)^{17}\text{C}$ recently measured at GANIL~\cite{LOISFUENTES25} focusing on the analysis of the $N=16$ shell gap.

The paper is organised as follows.
Sec.~\ref{sec2} introduces the formalism associated with deformed two-body models, in particular, considering the semimicroscopic NAMD model. 
Special emphasis is placed on the treatment of the continuum and the use of a pseudo-states basis.
Section~\ref{sec3} focuses on the implementation of the model and the pseudo-states discretisation method in transfer-to-the-continuum calculations.
The application of the formalism to the study of the transfer reaction $^{16}\text{C}(d,p)^{17}\text{C}$ is presented in Sec.~\ref{sec4}.
Finally, in Sec.~\ref{sec5} we summarise the main conclusions of this work.

\section{Deformed two-body models in a pseudo-states basis}\label{sec2}
The structure of the considered weakly bound nucleus is described using a two-body model composed of a weakly bound neutron and a deformed core. 
In particular, we employ the semimicroscopic Nilsson+AMD model (NAMD) recently proposed~\cite{Punta25,Tesis}.
In this model, the Hamiltonian of the system is written as 
\begin{equation}\label{eq:H}
{\cal H}=T(\vec r)+V_{\ell s}(r)(\vec \ell\cdot\vec s)+V_{vc}(\vec r,\xi)+h_{core}(\xi),
\end{equation}
where $T(\vec r)$ is the kinetic energy operator for the relative motion between the valence and the core.
$V_{vc}(\vec r,\xi)$ is the effective valence-core interaction,
which depends on the relative motion between the valence and the core,
but also on the core degrees of freedom $\xi$.
As explained in Ref.~\cite{Punta25}, this deformed potential is obtained convoluting the effective nucleon-nucleon interaction of Jeukenne, Lejeune, and Mahaux~\cite{JLM} with microscopic core densities calculated with Antysymmetrized Molecular Dynamics (AMD)~\cite{Kan13b}.
A spin-orbit term with the usual radial dependence $V_{\ell s}(r)$ is added to the valence-core interaction.
Finally, assuming that the core is a rigid rotor with a moment of inertia $\mathcal J$ and angular momentum $\vec I$, $h_{core}={\hbar^2\vec I^2}/{2\mathcal J}$.

The core+valence neutron models face the challenge of dealing with the Pauli exclusion principle, since the factorisation of the system does not allow complete antisymmetrization
In the NAMD model, Pauli-blocking effects related to the Nilsson states occupied by the $N$ core neutrons can be taken into account with different methods.
These effects are often ignored in the calculations; states with a dominant contribution from the lowest $N$ Nilsson orbitals are simply discarded. Such option is referred to here as  no-blocking method (NoB). Alternatively, we will consider a total blocking method (TB).
It consists of excluding from the calculation the $N$ Nilsson levels of lower energy~\cite{Punta25}.

In both cases, the constructed Hamiltonian $\mathcal H$ of the two-body model is diagonalised using a pseudo-states basis as explained in \cite{Punta23}. In particular, we employ the transformed harmonic oscillator basis (THO) with the analytical local scale transformation of Ref.~\cite{Kar05}.
This basis has been successfully applied to the structure and reactions of two-and three-body systems and has been generalised to the case in which core excitations are included \cite{Mor09,Lay12}.

The eigenvalues of the Hamiltonian are denoted as $\varepsilon_i^{J^\pi}$, and their corresponding wavefunctions can be written as
\begin{equation}\label{eq:thowf}
\Psi_{i M}^{J^\pi}(\vec{r},\xi)=\sum_{\alpha}
R^{J^\pi}_{i\alpha}(r)\Phi_{\alpha J}^{M}(\hat{r},\xi).
\end{equation}
They are characterised by the parity $\pi$, the total angular momentum of the nucleus $J$,
its $z$-projection $M$ and the number $i$ of the pseudo-state.
$\Phi_{\alpha J}^M(\hat{r},\xi)$ refers to the eigenstates of $J^2$ and $J_z$ 
resulting from the coupling of the angular momentum of the valence nucleon $\vec{j}$ to the core angular momentum $\vec{I}$,
\begin{equation}\label{eq:LabAng}
\Phi_{\alpha J}^M(\hat{r},\xi)\equiv\left[  
{\cal Y}_{\ell s}^j(\hat{r}) \otimes \phi_{I}(\xi) \right]_{JM}.
\end{equation}
In this expression, ${\cal Y}_{\ell s}^{jm}(\hat{r},\xi)$ denotes the wave function resulting from coupling the spin the valence particle $\vec s$ with the spherical harmonic corresponding to the orbital angular momentum $\vec{\ell}$ of the valence particle relative to the core.
The core is assumed to be a rigid rotor, so $\phi_{I}(\xi)$ is the eigenfunction of the Hamiltonian associated with the collective rotational motion of the core.
The label $\alpha$, which we call channel, denotes the set of quantum numbers $\{\ell,s,j,I\}$.

The interpretation of the pseudo-states depends on the sign of the eigenvalues.
The negative eigenvalues and their corresponding eigenfunctions are associated with the energies and wave functions of the bound states,
whereas the positive ones and their wave functions correspond to a discrete representation of the continuum.

The pseudo-states discretisation method can be very useful for some aspects of the continuum such as the description of resonances.
However, in some cases, such as in the evaluation of the energy differential cross sections discussed below, it is also necessary to calculate the \emph{exact} continuum wave functions $\Psi_{\alpha JM}^{(+)}(k_\alpha,\vec{r},\xi)$.
This wave function corresponds to an unbound state with a relative linear momentum $\hbar k_\alpha=\sqrt{2\mu_{nC}(\varepsilon-\mathcal E_I)}$ between the neutron and the core, where $\mu_{nC}$ is the reduced mass of the two-body system and $\mathcal E_I$ is the excitation energy of the core in the channel $\alpha$.
The sign $(+)$ refers to the outgoing boundary conditions explained later, and $\alpha$ is the incoming channel.  

The \emph{exact} continuum wave functions are calculated by solving the Schr\"odinger equation $\mathcal H\Psi_{\alpha JM}^{(+)}=\varepsilon\Psi_{\alpha JM}^{(+)}$ for positive energies $\varepsilon$, using the R-matrix method~\cite{Descouvemont10}.
The resulting wave functions can be written as
\begin{equation}\label{eq:cont_wf}
\Psi_{\alpha JM}^{(+)}(k_\alpha,\vec{r},\xi)=\sum_{\alpha'}
\frac{f^{J}_{\alpha\alpha'}(k_\alpha,r)}{r}\Phi_{\alpha' J}^{M}(\hat{r},\xi),
\end{equation}
where the radial functions $f^{J}_{\alpha\alpha'}(k_\alpha,r)$ are solutions of coupled differential equations,
\begin{equation}\label{eq:CoupledEq}
    \varepsilon f^{J}_{\alpha\alpha'}(k_\alpha,r)=
    \sum_{\alpha''} \langle\Phi_{\alpha' J}^{M}|{\cal H}|\Phi_{\alpha'' J}^{M}\rangle f^{J}_{\alpha\alpha''}(k_\alpha,r).
\end{equation}

The incoming channel $\alpha$ must be an open channel, that is, a channel for which $\varepsilon>\mathcal E_{I}$.
For a positive energy $\varepsilon$, there is always at least one open channel.
In case of channels $\alpha'$, we must consider all open channels and also the closed channels, for which $\varepsilon<\mathcal E_{I'}$.
These functions must behave asymptotically as follows,
\begin{equation}\label{eq:asymptotic}
     f^{J}_{\alpha\alpha'}(k_{\alpha},r)\xrightarrow[r\rightarrow\infty]{}
     \begin{cases}
     \frac{i}2
     \left[\delta_{\alpha\alpha'}H_{\ell'}^{(-)}(k_{\alpha'},r)-
     S^J_{\alpha\alpha'}H_{\ell'}^{(+)}(k_{\alpha'},r)\right] & \varepsilon>\mathcal E_{C'} \\
     A_{\alpha\alpha'}W_{l'+\frac12}(-2ik_{\alpha'}r) & \varepsilon<\mathcal E_{C'}
          \end{cases},
\end{equation}
where $H_{\ell'}^{(-)}(k_{\alpha'},r)$ and $H_{\ell'}^{(+)}(k_{\alpha'},r)$ are incoming and outgoing Coulomb spherical wave functions respectively, and $W_{l'+\frac12}(-2ik'_\alpha r)$ are the Whittaker functions.
$S^J_{\alpha\alpha'}$ are elements of the S-matrix and $A_{\alpha\alpha'}$ are the asymptotic normalisation constants.

We introduce the energy density of the pseudo-states, defined from their overlap  with the \emph{exact} continuum wave functions as follows:
\begin{equation}\label{eq:Goverlap}
     \mathcal{G}_{\alpha i}^J(k_\alpha)\equiv
     \langle \Psi_{\alpha JM}^{(-)}|\Psi_{iM}^{J^\pi}\rangle=
     \sum_{\alpha'} \int f^{J}_{\alpha\alpha'}(k_{\alpha},r)R^{J^\pi}_{i\alpha'}(r)rdr,
\end{equation}
where $\Psi_{A}^{(-)}$ is just the time reversal of $\Psi_{A}^{(+)}$.
Therefore, they are defined as in Eq.~(\ref{eq:cont_wf}) but with conjugated radial functions $f^{J*}_{\alpha\alpha'}(k_{\alpha},r)$. These densities are needed to recover a continuous distribution of the transfer to the continuum from the discrete calculations obtained with the pseudo-states.

\section{Transfer to the continuum using pseudo-states}\label{sec3}
We consider a transfer reaction of the form $C(d,p)A$, where $A$ is a weakly bound nucleus studied with our deformed two-body model and $C$ its corresponding core.
In this process, the deuteron collides with the nucleus $C$, both in their ground state, with angular momenta $J_d$ and $I_0$ respectively.
After the collision, a proton and a nucleus $A$ emerge, allowing different states to be populated, with different angular momentum and parity $J^\pi$, for the latter.  
 To describe this reaction, we combine the finite-range adiabatic distorted-wave approximation (ADWA)  reaction framework ~\cite{JT}  with the recently developed NAMD structure model. 
The NAMD model has already been implemented in ADWA calculations for transfer reactions populating bound states, with encouraging results \cite{Punta25}. 
However, the formalism has to be adapted for the transfer to unbound states in the continuum.

In the case of transfer to the continuum, the nucleus $A$ spontaneously breaks into the system $C+n$, with quantum numbers $\alpha\equiv\{\ell,s,j,I\}$.
We can calculate the cross section associated with this channel $\alpha$, when unbound states of $A$ are populated, with angular momentum and parity $J^\pi$.
Since there is a continuum of unbound states, energy distributions are considered.
In particular, two-body energy-angular cross section distributions can be calculated as follows: 
\begin{equation}\label{dxsde}
    \frac{d^2\sigma_{C(d,p)A}}{d\Omega_{pA}dE} =
    \frac{2\mu_{nC}}{\pi\hbar^2k_\alpha}
    \frac1{(2J_d+1)(2I_0+1)}
    \sum_{\zeta} \left|\mathcal{F}_{C(d,p)A}\right|^2.
\end{equation}
The energy $E$ is considered with respect to the neutron separation threshold, which corresponds to the eigenvalue of the Hamiltonian $\mathcal{H}$.
$\Omega_{pA}$ refers to the orientation of the relative momentum $\hbar\vec{k}_{pA}$ for the system $p+A$.
As in the previous section, $\hbar k_\alpha$ corresponds to the relative momentum of the $C+n$ system, and $\mu_{nC}$ to its reduce mass.
In Eq.~(\ref{dxsde}), assuming that the projections of angular momentum are not distinguished; we sum over the possible final projections, and the average is considered over all the possible projections for the initial fragments.
Then, $\zeta$ denotes the set of all these projections.
The scattering amplitude $\mathcal{F}_{C(d,p)A}$ is defined as
\begin{equation}\label{Fdef}
\mathcal{F}_{C(d,p)A}=\sqrt{\frac{\mu_{dC}\mu_{pA}k_{pA}}{(2\pi\hbar^2)^2k_{dC}}}
 \langle\chi_{\vec k_{pA}}^{(-)}\varphi_p\Psi_{A}
 |\Delta U|\chi_{\vec k_{dC}}^{(+)}\psi_{d}\phi_{C}\rangle.
\end{equation}
Here, $\chi_{\vec k_{dC}}^{(+)}$ and $\chi_{\vec k_{pA}}^{(-)}$ are respectively distorted waves for the entrance and exit channels, dependent on the relative momentum of the initial system $d+C$, and final $p+A$.
$\mu_{dC}$ and $\mu_{pA}$ denote, respectively, the reduced masses of the systems.
The function $\psi_d$ correspond to deuteron wave function and $\varphi_{p}$ represents the state of the emerging proton (the spin state).
Considering the NAMD model, the state of the nucleus $A$ is described by the function $\Psi_A\equiv\Psi_{\alpha JM}(k_{\alpha},\vec r,\xi)$.
In the same way, $\phi_C\equiv\phi_{I_0}^{M_0}(\xi)$ describe the state of the nucleus $C$.
Note that expressions (\ref{dxsde}) and (\ref{Fdef}) have been obtained assuming normalisation 
$\langle\chi_{\vec k}|\chi_{\vec k'}\rangle=(2\pi)^3\delta(\vec k-\vec k')$ for all the scattering states.

Calculating the scattering amplitude for the continuum of unbound states is not feasible in general, and we have to discretise the continuum.
The pseudo-states method is an efficient way to discretise the continuum, which has already proven to be very successful in its application to breakup reactions~\cite{deD17}.
To implement this method in transfer-to-the-continuum calculations, we define the following scattering amplitudes:
\begin{equation}\label{Fmatrix_i}
\mathcal{F}^{J^\pi}_{i\zeta}(\hat k^i_{pA})=
\sqrt{\frac{\mu_{dC}\mu_{pA}k_{pA}}{(2\pi\hbar^2)^2k_{dC}}}
\langle\chi_{\vec k^i_{pA}}^{(-)}\varphi_p\Psi_{iM}^{J^\pi}
|\Delta U|\chi_{\vec k_{dC}}^{(+)}\psi_{d}\phi_{I_0}^{M_0}\rangle,
\end{equation}
which are associated with the pseudo-states $\Psi_{iM}^{J^\pi}$ obtained with our structure models.
The momentum $\hbar k_{pA}^i$ is associated with $E_{pA}^i=E_{dC}-S_n(d)-\varepsilon_i^{J^\pi}$, where $S_n(d)$ is the neutron separation energy of the deuteron.

The scattering amplitudes of Eq.~(\ref{Fmatrix_i}) are calculated using FRESCO code~\cite{fresco}.
In these calculations, the \emph{prior} form of the transition potential, $\Delta U=V_{nC}+U_{pC}-U_{dC}$, is used.
Note that $V_{nC}$ is the two-body interaction of the Hamiltonian $\mathcal H$ corresponding to our structure model for the nucleus $A$.
Then, the calculation of the scattering matrix depends on the so-called vertex functions,
\begin{equation}\label{overlaps_vertex}
\langle\Psi_{iM}^{J^\pi}|V_{nC}|\phi_{I_0}^{M_0}\rangle=
\sum_{jm}\langle jmI_0M_0|JM\rangle[\mathcal V_{i\alpha}^{J^\pi}(r){\cal Y}_{\ell s}^{jm}(\hat{r})]^\dagger,
\end{equation}
where the radial functions $\mathcal V_{i\alpha}^{J^\pi}(r)\equiv\langle\Phi_{\alpha M}^J|V_{nC}|\Psi_{iM}^{J^\pi}\rangle$ are extracted from our deformed two-body models.
The channel $\alpha_0$ corresponds to the channel $\alpha$ with $I=I_0$ and $\langle jmI_0M_0|JM\rangle$ is a Clebsch-Gordan coefficient.

We assume that the \emph{remnant} of the transition potential ($U_{pC}-U_{dC}$), only depends on the relative coordinates $\vec r_{pC}$ and $\vec r_{dC}$. 
Consequently, its contribution to the scattering amplitude depends on the structure of the nucleus $A$ through the overlap,
\begin{equation}\label{overlap}
\langle\Psi^{J^\pi}_{i M}|\phi^{M_0}_{I_0}\rangle=
\sum_{j}\langle jmI_0M_0|JM\rangle R_{i\alpha_0}^{J^\pi}(r)\left[{\cal Y}_{\ell s}^{jm}(\hat{r})\right]^\dagger.
\end{equation}
These overlaps are the only required ones in the \emph{post} form; vertex functions are not involved.
However, in transfer-to-the-continuum calculations, numerical problems arise when using the \emph{post} form.
These problems are associated with slow convergence in the calculations because the radial functions $R_{i\alpha_0}^{J^\pi}(r)$ of unbound states do not vanish rapidly.
The convergence problems are avoided using the \emph{prior} form because in this case the scattering amplitude are mostly determined by the vertex functions of Eq.~(\ref{overlaps_vertex}).
Note that due to the action of the potential, the functions $\mathcal V_{i\alpha}^{J^\pi}(r)$ vanish at very short distances. Further details can be found in~\cite{Tesis}.

Once the discrete scattering amplitudes $\mathcal{F}^{J^\pi}_{i\zeta}(\hat k^i_{pA})$ have been calculated, the continuous energy distribution of the cross section can be obtained by assuming
\begin{equation}\label{Fmatrix_conv}
    |\mathcal{F}_{C(d,p)A}(k_\alpha,\hat k_{pA})|^2 \approx
    \left|\sum_i\mathcal G_{\alpha i}^J(k_\alpha)
    \mathcal{F}_{i\zeta}(\hat k^i_{pA})\right|^2. 
\end{equation}
This expression is based on the completeness relation of the basis, which is approximately valid if a sufficiently large number of pseudo-states is taken for the considered energy range. Convergence of the calculated cross-section with respect to the number of pseudo-states has been checked for this purpose.

\section{Application to $^{16}\text{C}(d,p)^{17}\text{C}$}\label{sec4}
The transfer reaction $^{16}\text{C}(d,p)^{17}\text{C}$ has recently been studied~\cite{LOISFUENTES25}, with the aim of localising the $1d_{3/2}$ orbital for neutrons in $^{16-17}$C, to confirm the $N=16$ shell gap.
Once the NAMD model has been tested in the case of transfer to the bound states of $^{17}$C~\cite{Punta25}, we now extend the application to the transfer to unbound states using the formalism previously introduced.

Transfer-to-the continuum calculations are performed using the same interactions $V_{pn}$,$U_{pA}$, and $U_{dC}$ as in Refs.~\cite{Punta23,Punta25}.
Experimental data was obtained in GANIL in inverse kinematics, using a $^{16}\text{C}$ beam at 17.2~MeV/nucleon and measuring the protons emitted at the most backward angles in the laboratory frame ($147^\circ\le\theta_{lab}\le167^\circ$)~\cite{Pereira,LOISFUENTES25}.
This corresponds to very small angles for $^{17}$C in the center-of-mass frame ($\theta_{CM}<11^\circ$), although the exact range depends on the energy of the populated state.

\begin{figure}
	\centering 
	\includegraphics[width=0.45\textwidth]{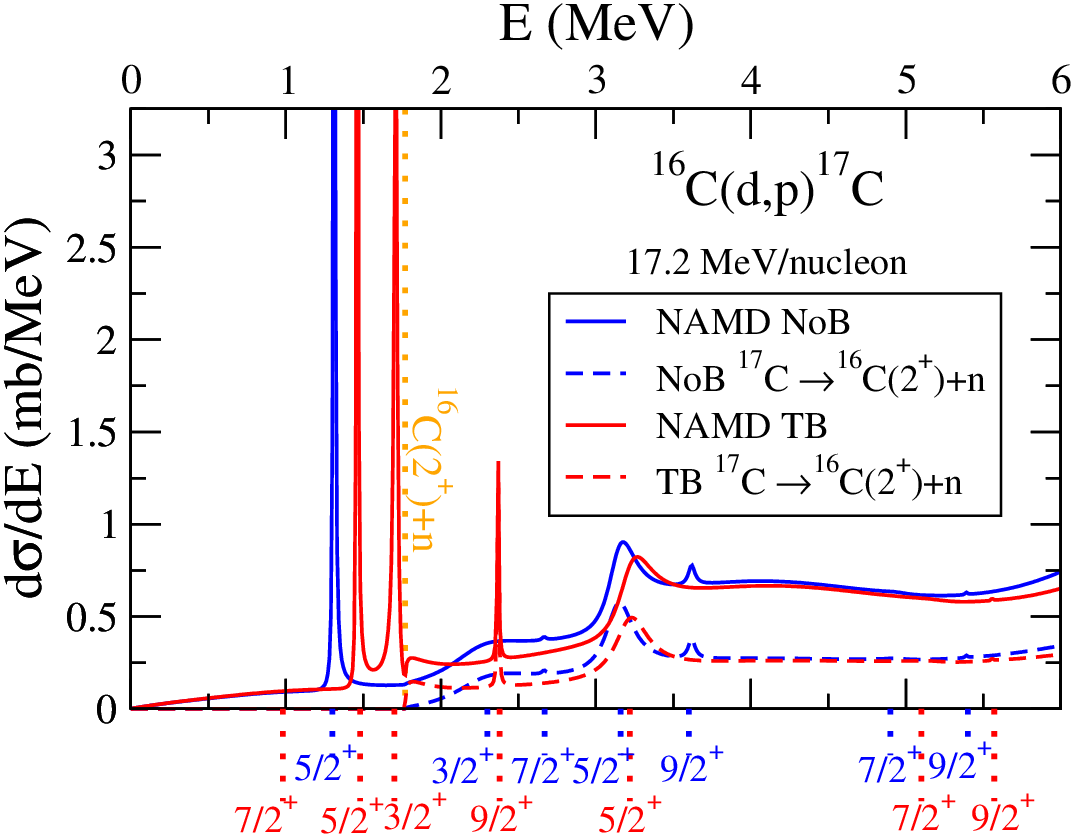}	
\caption{Energy distribution with respect to the neutron separation threshold of the transfer to the continuum for the reaction $^{16}$C$(d,p)^{17}$C at 17.2~MeV/nucleon.
Calculations have been made using the NAMD model with the NoB and TB methods, for which the predicted resonances are marked with dotted lines.
The sum of all contributions (solid lines) are shown along with the distributions
when the $^{16}\text{C}$ nucleus results in its first excited state $2^+$ (dashed lines).} 
	\label{c16dpc17_spectrum}%
\end{figure}
We calculate the energy distributions of the cross section for the experimental angular range, using the NAMD model with NoB and TB methods~\cite{Punta25}.
To describe the $^{17}$C continuum, we consider pseudo-states up to 13~MeV above the neutron separation threshold, with $J\le9/2$ in the case of positive parity and $J\le7/2$ in the case of negative parity.
The energy distributions of the total cross section, that is, the sum of all contributions, are shown in Fig.~\ref{c16dpc17_spectrum} with solid lines.
The dashed lines correspond to the sums of the contributions for the outgoing channels with the core state $2^+$.
Note that, since unbound states of $^{17}$C are populated, the nucleus immediately emits a neutron, and the distributions mentioned correspond to the case where $^{16}$C results in its first excited state $2^+$.
This distribution was measured experimentally in addition to that of the total cross section using particle-gamma coincidences~\cite{LOISFUENTES25}.

On the $x$-axis, the resonances predicted using the NAMD model with the NoB and TB methods are marked.
Several resonances of $^{17}$C are predicted, but many of them weakly contribute to the transfer cross section because they are strongly associated with a $d_{5/2}$ configuration for the valence neutron coupled to excited states of the core.
It must be taken into account that the transfer reaction $^{16}\text{C}(d,p)^{17}\text{C}$ starts with $^{16}$C in its ground state $0^+$, which reduces the probability of populating states of $^{17}$C with minimum weights for the component associated with this core state.

\begin{figure}
	\centering 
	\includegraphics[width=0.45\textwidth]{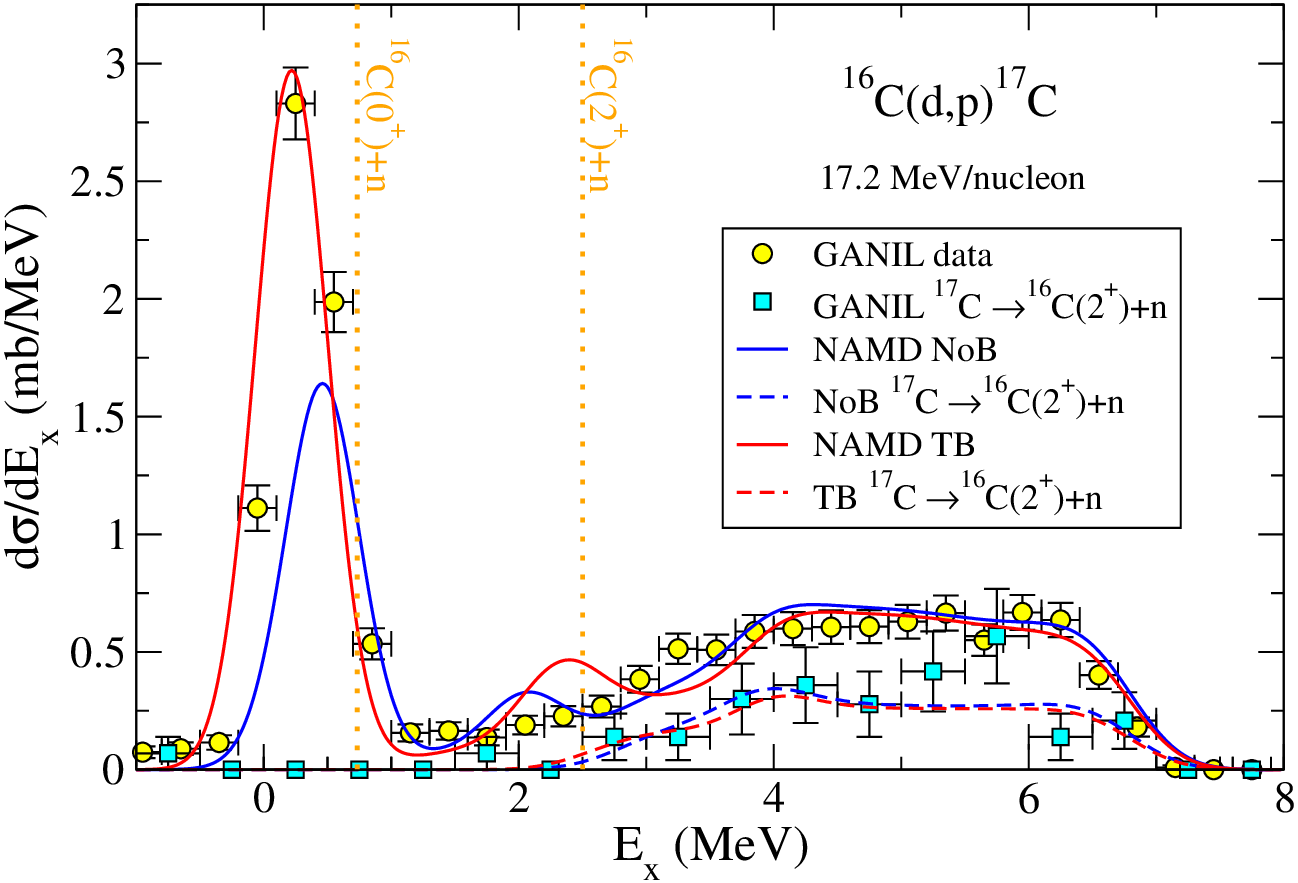}	
\caption{Excitation energy distribution of the cross section for the transfer reaction $^{16}$C$(d,p)^{17}$C at 17.2~MeV/nucleon obtained using the NAMD model with the NoB and TB methods.
The distributions of the total cross section (solid line) and that associated with the $^{16}\text{C}(2^+)+n$ decay of $^{17}$C (dashed line) are compared with the experimental data from GANIL~\cite{LOISFUENTES25}.} 
	\label{c16dpc17}%
\end{figure}
In order to compare our calculations with the experimental data, we convolute the theoretical distributions with the experimental energy resolution.
The results are shown in Fig.~\ref{c16dpc17}, also including the contribution of the transfer to the bound states, and they are compared with the experimental data from GANIL \cite{LOISFUENTES25}.
Note that, for a meaningful comparison with the experimental data,  before convolution we have transformed the energy with respect to the neutron separation threshold to the excitation energy by adding the one neutron separation energy of $^{17}$C.

As discussed in Ref.~\cite{Punta25}, the transfer to the bound states is underestimated when Pauli blocking is not taken into account (i.e., the NoB method), while it is well reproduced when Pauli blocking is imposed using the TB method.
Meanwhile, using both methods, the agreement with the data for the transfer to the continuum is similar.
In the case of TB, the contributions of the first $3/2^+$ and $5/2^+$  resonances result in a peak around 2.5~MeV, that is not consistent with the data.
Note that these resonances are very close to the $^{16}\text{C}(2^+)+n$ threshold, so small differences in energy can have strong effects in the reaction.
In any case, the general agreement of the calculations using the NAMD model considering the TB is fairly good, taking into account the complexity of the problem and the simplicity of the model.
For the distribution associated with $^{16}\text{C}(2^+)+n$ decay of $^{17}$C, good agreement is found using both methods.
The figure also shows that the results for the total cross section obtained using different blocking methods become more similar as the energy increases. This can be understood by noting that the single-particle states occupied by the core nucleons are those with the lowest energies and, therefore, the blocking must affect more strongly the lower part of the spectrum.

\begin{figure}
	\centering 
	\includegraphics[width=0.45\textwidth]{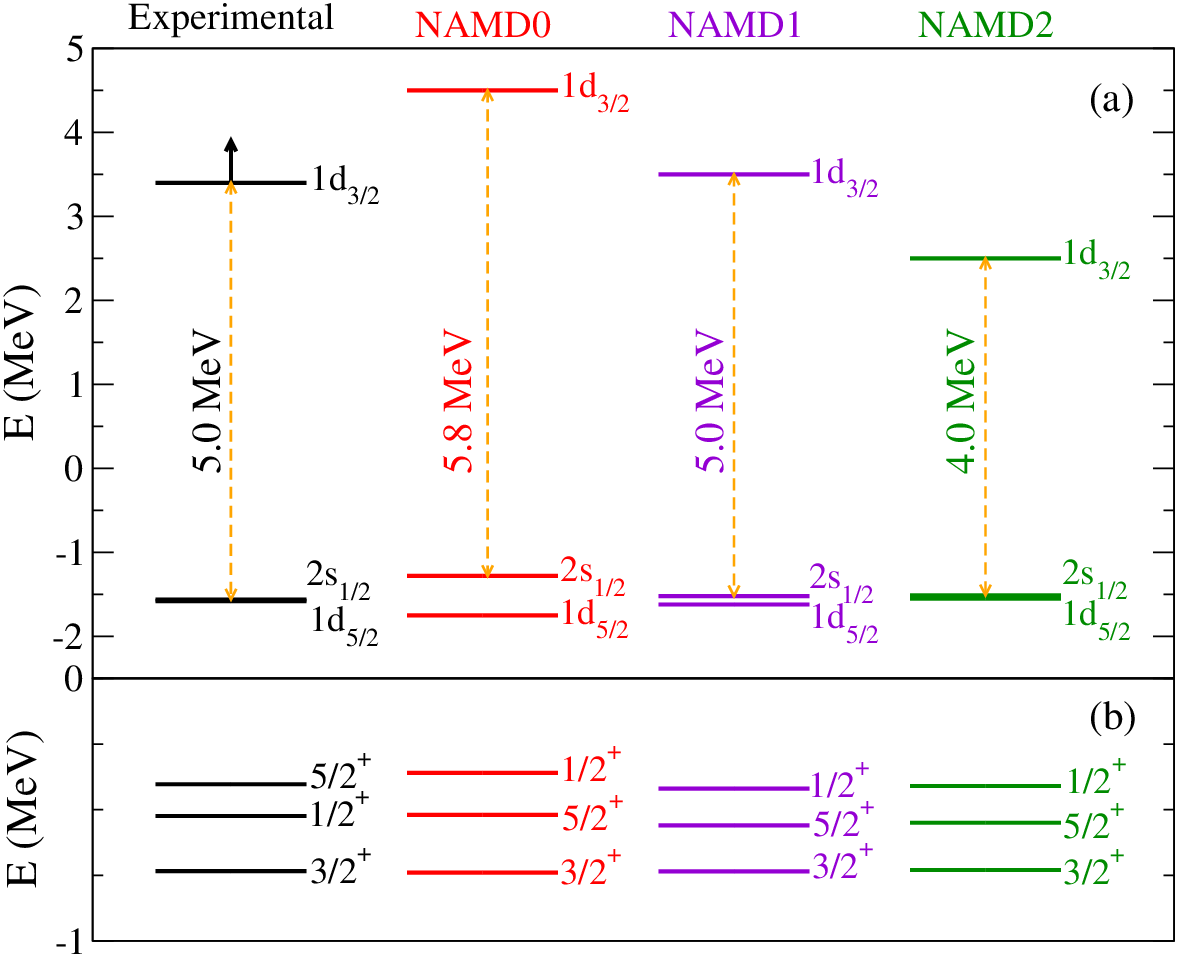}	
\caption{Energy levels obtained using different variations of the NAMD model with the TB method. In the upper panel, the spherical single-particle levels are compared with those extracted from GANIL data~\cite{Pereira,LOISFUENTES25}.
The arrow marks that the energy showed for the level $1d_{3/2}$ is a lower limit.
In the lower panel, the energies of the bound states of $^{17}$C are compared with the corresponding experimental values~\cite{Wang21,Ele05}.} 
	\label{levels_vs}%
\end{figure}

A key point of this reaction is that the transfer to the continuum begins to be significant from approximately 2~MeV above the neutron separation threshold.
In this transfer process, the population of $3/2^+$ states strongly associated with the  $1d_{3/2}$ orbital stands out.
The energy separation of these states from the bound states of $^{17}$C suggests the existence of the $N=16$ shell gap between the $1d_{3/2}$ and the $2s_{1/2}$ orbitals.
We study the effect of this shell gap, $\Delta$, in the calculated cross section using different variations of the NAMD model.
From the analysis of the transfer data, a lower limit $\Delta=5.08^{+0.43}_{-0.33}$~MeV was estimated for the shell gap~\cite{LOISFUENTES25}.
The already presented NAMD model with total blocking entails a $N=16$ shell gap of $\Delta=5.8$~MeV, larger than this value.
Therefore, considering also the TB method, we have tested two variants reducing the spin-orbit splitting: One with $\Delta=5.0$~MeV and other with $\Delta=4.0$~MeV.
In Fig.~\ref{levels_vs}(a), the spherical single-particle levels obtained with these variants of the NAMD model are compared with those proposed in \cite{Pereira,LOISFUENTES25}.
It can be seen that they differ mainly in the energy of the $1d_{3/2}$ orbital.
Fig.~\ref{levels_vs}(b) compares the energies of the bound states of $^{17}$C obtained using the three variants of the NAMD model with the experimental energies~\cite{Wang21,Ele05}, showing that the variants weakly affect the energies of these states.
\ref{appendix} provides further details on the specific parameterisation of the models.

\begin{figure}
	\centering 
	\includegraphics[width=0.45\textwidth]{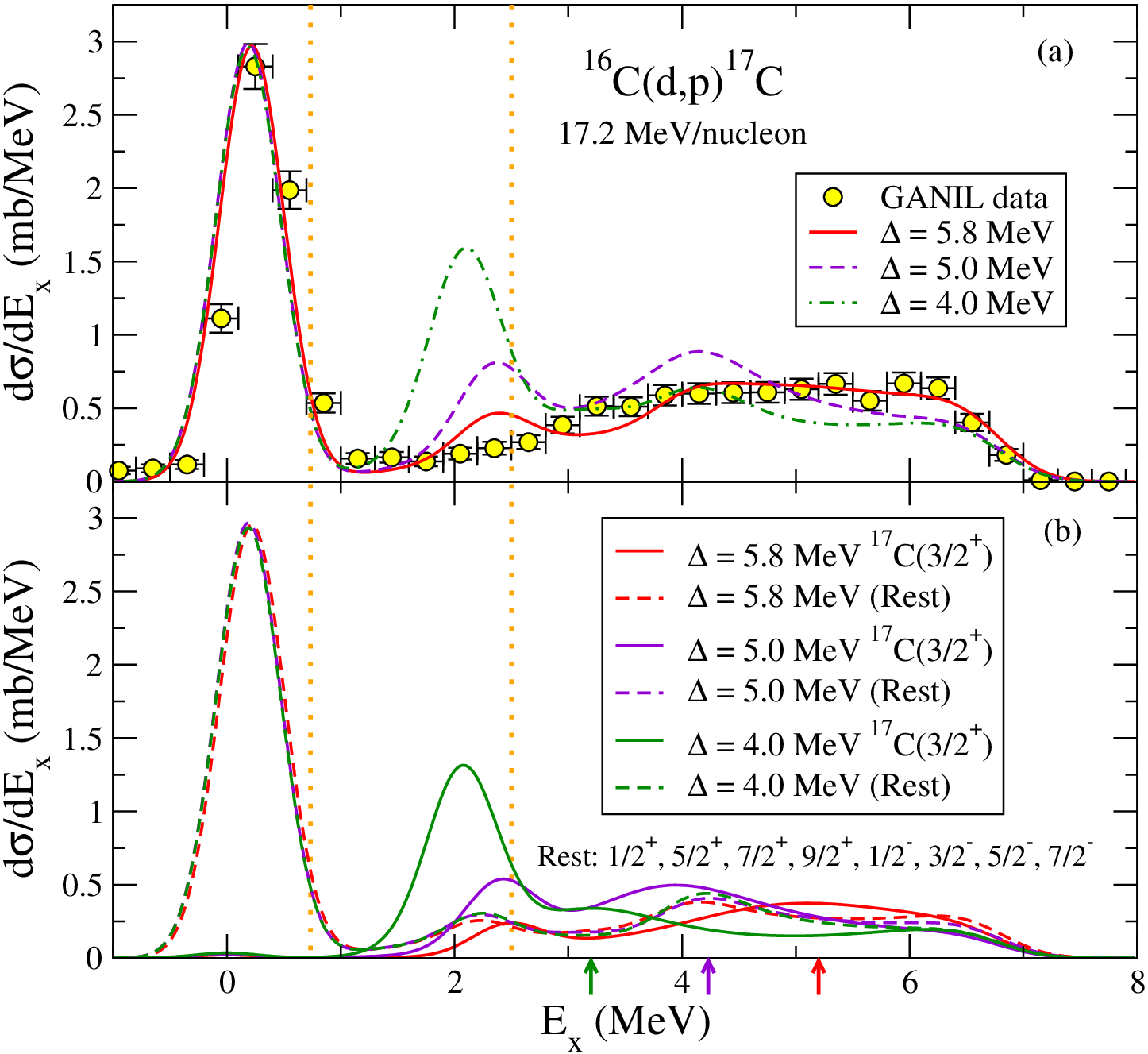}	
\caption{Excitation energy distribution of the total cross section for the transfer reaction $^{16}$C$(d,p)^{17}$C obtained using variations of the NAMD model with different values of the $N=16$ shell gap $\Delta$.
The distributions of the total cross section are compared with the experimental data from GANIL~\cite{LOISFUENTES25} (upper panel).
The distributions when $3/2^+$ states of $^{17}$C are populated are isolated from the sums of the rest of contributions in the lower panel.
The energy of the orbital $1d_{3/2}$ for each calculation is marked with an arrow on the x-axis.} 
	\label{c16dpc17_vs}%
\end{figure}
The results of the transfer calculations using these models are shown in Fig.~\ref{c16dpc17_vs}.
In the upper panel, the obtained energy distributions of the total cross section are compared with the experimental data~\cite{LOISFUENTES25}.
It is clear that the discrepancy with the data increases when the spin-orbit splitting is reduced.
In particular, the peak around 2~MeV becomes larger, and the continuous contribution between about 3 and 6~MeV is more concentrated at the lower energies.
This is due to the change in the contribution associated with the population of $3/2^+$ states.
These distributions and the sum of the remaining contributions are shown in the lower panel of Fig.~\ref{c16dpc17_vs} for the three models.

It can be seen that the distribution associated with the $3/2^+$ states is strongly affected by shifting the $1d_{3/2}$ level, while the rest of the contributions remain almost invariant.
As expected, by decreasing the energy of the spherical level, the strength is distributed to lower energies.
When the shell gap is reduced, the energy of the first $3/2^+$ resonance  decreases and its contributions increase significantly, thus increasing the discrepancy with the data.
As already discussed, this contribution is very sensitive to small differences in energy, and the discrepancies could be partially due to limitations of the NAMD model.
However, it is clear that, in order to concentrate the cross section in the region between about 3 and 6~MeV as the data show, the energy of the $1d_{3/2}$ level should be greater than 3~MeV above the neutron separation threshold, leading to a shell gap $N=16$ larger than $5$~MeV.
Therefore, the results obtained with the NAMD model corroborate the lower limit extracted from the previous analysis of the experimental data ($\Delta=5.08^{+0.43}_{-0.33}$~MeV)~\cite{LOISFUENTES25}.

\section{Conclusions}\label{sec5}
In summary, we have presented a new framework for studying transfer to the continuum reactions by describing the unbound states within deformed two-body models using pseudo-states.
This formalism has allowed us to study the transfer reaction $^{16}\text{C}(d,p)^{17}\text{C}$ for which experimental data are available.

In our calculations, the recently developed NAMD model is employed to describe the $^{17}$C structure, considering the no-blocking (NoB) and total blocking (TB) methods for the treatment of Pauli-blocking effects.
The NAMD model predicts several resonances for $^{17}$C that weakly contribute to the transfer cross section because they are mainly associated with excited states of the core.
On the other hand, the population of $3/2^+$ states, strongly associated with a neutron in the spherical orbital $1d_{3/2}$ and the core state $0^+$, plays a key role.

Using both the NoB and the TB methods, the resulting energy distributions of the cross section are in good agreement with the experimental data for the transfer to the unbound states.
However, the transfer to the bound states is clearly better reproduced using the TB method, highlighting the importance of Pauli-blocking effects.
In addition, our calculations support the hypothesis of a large $N=16$ shell gap in this region ($^{16-17}$C).
In particular, the NAMD model with TB presents a shell gap of 5.8~MeV. Nevertheless, we take advantage of the flexibility of the present description to test different variations of the model reducing the shell gap by placing the $1d_{3/2}$ orbital at lower energies.
From the results of these calculations we extract that the $N=16$ shell-gap must be larger than 5~MeV to adequately describe the experimental data, corroborating what was suggested in  Ref.~\cite{LOISFUENTES25}.

In view of the results, the NAMD model with TB provides a good description of $^{17}$C in its application to the transfer reaction $^{16}\text{C}(d,p)^{17}\text{C}$. 
As a next step, it is important to check that the model is also suitable for describing other reactions involving $^{17}$C.
For example, the application to the breakup reaction $^{17}\text{C}(p,p')^{16}\text{C}+n$ is already being studied. 

In addition, the formalism presented here combines flexibility with a proper treatment of relevant degrees of freedom. Therefore, it can be very useful for studying other transfer reactions populating unbound states of other exotic nuclei, such as $^{18}\text{C}(d,p)^{19}\text{C}$ and $^{30}\text{Ne}(d,p)^{31}\text{Ne}$ and, in general, to help in the study of shell evolution including unbound states in the different islands of inversion.

\bigskip
\section*{Acknowledgements}   
The present research is funded from grant PID2023-146401NB-I00 by MICIU/AEI/10.13039/501100011033 and by FEDER, UE.
P.P. acknowledges PhD grant FPU21/03931 from the Ministerio de Ciencia, Innovaci\'on y Universidades.

\appendix
\section{}\label{appendix}
The variations of the NAMD model of Fig.~\ref{c16dpc17_vs} have been obtained by reducing the spin-orbit splitting with respect to the NAMD model with total blocking (TB) presented in Ref.~\cite{Punta25}. 
We have considered a spin-orbit potential parametrised in terms of the derivative of the Wood-Saxon function ($f_{WS}$),
\begin{equation}\label{eq:prm_ls}
    V_{\ell s}(r)=\frac{4V_{so}}r\frac{df_{WS}(r)}{dr}.
\end{equation}
The geometry is extracted from the fitting of the central potential of the NAMD model: $R=2.9$~fm $a=0.7$~fm.
$V_{so}=8$~MeV is used to obtain $\Delta=5$~MeV, and $V_{so}=6$~MeV for $\Delta=4$~MeV. 

A slight renormalisation of the potential is also applied for $l=2$ ($1.01$ for $\Delta=5$~MeV and $1.04$ for $\Delta=4$~MeV), to maintain the energy of the $1d_{5/2}$ level.
In turn, the global renormalisation of the central potential is corrected to obtain a ground-state energy consistent with the experimental value ($1.019$ for $\Delta=5$~MeV and $1.018$ for $\Delta=4$~MeV).

\bibliography{tesis}

\end{document}